\documentclass[useAMS,usenatbib]{mnras}
\usepackage{graphicx}
\usepackage{epstopdf,color,verbatim}
\usepackage{threeparttable}
\usepackage[]{txfonts}
\usepackage{caption}
\usepackage{threeparttable}
\usepackage{multicol}

\def\simless{\mathbin{\lower 3pt\hbox
{$\rlap{\raise 5pt\hbox{$\char'074$}}\mathchar"7218$}}}   
\def\simmore{\mathbin{\lower 3pt\hbox
{$\rlap{\raise 5pt\hbox{$\char'076$}}\mathchar"7218$}}}   
\def\Msun{{\rm M}_\odot}                                       
\newcommand{\bi}{\begin{itemize}}
\newcommand{\ei}{\end{itemize}}

\usepackage{epsfig} 
\usepackage{xcolor}
\usepackage{epstopdf}
\title[]
{Shocks and dust formation in Nova V809 Cep}

\author[A-N Babul et al.]{
Aliya-Nur Babul,$^{1}$\thanks{E-mail:aliya.babul@columbia.edu}
Jennifer L. Sokoloski,$^{1}$
Laura \ Chomiuk,$^{2}$
Justin D. Linford,$^{3}$ 
\newauthor
Jennifer H.S. Weston,$^{4}$
Elias Aydi,$^{2}$
Kirill V. Sokolovsky $^{2,5}$
Adam M. Kawash $^{2}$
\\
$^{1}$Department of Astronomy, Columbia University, 550 W 120th St, New York, NY 10027, USA\\
$^{2}$Center for Data Intensive and Time Domain Astronomy, Department of Physics and Astronomy, Michigan State University, East Lansing, MI 48824, USA\\
$^{3}$National Radio Astronomy Observatory, P.O. Box O, Socorro, NM 87801, USA\\
$^{4}$ Federated IT, 1201 Wilson Blvd, 27th Floor, Arlington VA 22209 \\
$^{5}$ Sternberg Astronomical Institute, Moscow State University, Universitetskii~pr.~13, 119992~Moscow, Russia
}

\begin{document}
\pagerange{\pageref{firstpage}--\pageref{lastpage}} \pubyear{2018}

\maketitle

\label{firstpage}

\begin{abstract}
     The discovery that many classical novae produce detectable GeV $\gamma$-ray emission has raised the question of the role of shocks in nova eruptions. Here we use radio observations of nova V809 Cep (Nova Cep 2013) with the Jansky Very Large Array to show that it produced non-thermal emission indicative of particle acceleration in strong shocks for more than a month starting about six weeks into the eruption, quasi-simultaneous with the production of dust. Broadly speaking, the radio emission at late times -- more than a six months or so into the eruption -- is consistent with thermal emission from $10^{-4}\,\Msun$ of freely expanding, $10^4$~K ejecta. At 4.6 and 7.4 GHz, however, the radio light-curves display an initial early-time  peak 76 days after the discovery of the eruption in the optical ($t_0$). The brightness temperature at 4.6 GHz on day 76 was greater than  $10^5 K$, an order of magnitude above what is expected for thermal emission. We argue that the brightness temperature is the result of synchrotron emission due to internal shocks within the ejecta. The evolution of the radio spectrum was consistent with synchrotron emission that peaked at high frequencies before low frequencies, suggesting that the synchrotron from the shock was initially subject to free-free absorption by optically thick ionized material in front of the shock.  Dust formation began around day 37, and we suggest that internal shocks in the ejecta were established prior to dust formation and caused the nucleation of dust. 
\end{abstract}

\section{Introduction}
A classical nova is an eruption that occurs in an interacting stellar binary system in which one star is a white dwarf and the other is typically a main sequence star \citep[][]{laura_review,della}. Hydrogen-rich material from the donor star accretes onto the white dwarf and is compressed, causing the temperature to rise. The soaring temperatures trigger a thermonuclear runaway and a subsequent ejection of matter from the white dwarf \citep[][]{starrfield}. As material is ejected from the white dwarf, the system brightens across much of the electromagnetic spectrum, allowing for the discovery of such phenomenon. 

Between its launch in 2008 June and 2021 April, 
the Large Area Telescope (LAT; \citealt{2009ApJ...697.1071A}) of the \emph{Fermi Gamma Ray Space Telescope} 
detected GeV gamma-rays from 17 
novae (see the list in \citealt{2020arXiv201108751C,fran}, plus the recurrent nova V3890~Sgr; \citealt{2019ATel13114....1B}), typically within a few days of their optical peak \citep[e.g.,][]{fermi2014,laura_review,2020arXiv201108751C}
The presence of GeV gamma-rays implies that these eruptions have generated a population of relativistic particles. 
These particles are likely produced via the diffusive shock acceleration mechanism as 
a result of internal shocks within the 
ejecta
\citep[e.g.,][]{1994ApJS...90..515B,2016MNRAS.457.1786M}. 
The high-energy particles may produce gamma-rays via a number of processes
including pion decay, inverse Compton scattering of ambient optical photons and relativistic bremsstrahlung 
\citep{2015MNRAS.450.2739M,2018A&A...612A..38M}. High-energy particles may also produce synchrotron radiation in the radio band \citep{vlasov}.
Radio imaging of nova V959~Mon revealed regions of synchrotron emission, further confirming the presence of shocks and providing a localization within the ejecta. From the images, \citet{chomui2014} postulated 
that the shocks are the result of 
collisions between two outflows from the white dwarf: 
the first, 
more slowly moving outflow having
a toroidal geometry, and the second, 
faster 
flow having a more spherical geometry. 
The shocks 
were located near the equatorial plane, where the two outflows collided. Based on the optical spectral evolution of a large sample of novae, \cite{2020ApJ...905...62A} found consistent evidence for this colliding outflows scenario
Finally, observations of correlated variations in GeV and optical brightness suggest
that shocks may transport a significant fraction of the nova eruption energy
\citep{2017NatAs...1..697L,v906} and power a substantial fraction of the bolometric luminosity during the early weeks of a nova eruption. 

In nova-producing binaries that are too distant for $\gamma$-ray detections, radio observations provide a unique opportunity to study developing shocks. Radio lightcurves sometimes
contain local maxima, which we refer to as {\em early-time flares}, within months of the initial optical peak \citep[][]{QUvul,v1324,Westonaql}. These early-time flares correspond to radio surface brightnesses (parameterized as brightness temperatures) that are at least an order of magnitude higher than 
expected from thermally emitting, $10^{4}$ K ionized gas that is expanding freely with speeds inferred from optical spectral lines \citep[][]{2015ApJ...803...76C}. Because the brightness temperature represents a lower limit on the temperature of thermally emitting material,
brightness temperatures in excess of $10^5$\,K indicate that either the 
the ejecta contain enough
T $\geq 10^{5} $ K gas for that gas to be optically thick to free-free emission at radio wavelengths or 
the radio emission is non-thermal emission 
from electrons accelerated via shocks. In at least one case (V1723~Aql), \cite{Westonaql} argued that the 
mass of $10^5$~K gas required for the ejecta to generate the observed radio emission through thermal processes
was unfeasible,
and that the high brightness temperature 
was more likely to be the result of synchrotron emission produced near the internal shocks in the ejecta \citep[][]{vlasov}.

Optical and Infared (IR) light curves of classical novae sometimes display characteristics of dust formation days to months after the initial eruption \citep[e.g.,][]{dust_novae,dust_2}. Sudden decreases in the optical flux are seen in about 20 percent of novae within 100 days of the optical peak \citep{strobe}. These decreases occur earlier than what would be expected due to the natural fading in nova luminosity over time, and -- along with the subsequent optical rebrightenings -- are referred to as ``dust dips''.  They 
result from the obscuring of some optical light from the nova due to the formation of dust and also 
often appear along with brightening at infrared wavelengths, a typical signature of dust emission \citep[][]{gehrz_2008}.

Dust production requires cool (T $\lesssim 10^3$~K) gas for the condensation of dust grains, and regions of dust condensation must be shielded from the ionizing radiation emanating from the white dwarf. \citet{andrea,dustform} reconciles the presence of harsh radiation with the onset of dust formation. Their model, which we refer to as the \emph{Shock-Dust Model}, postulates that dense radiative shocks, such as the ones thought to give rise to the $\gamma$-rays and early-time radio peaks, provide the optimal conditions for dust condensation. As both the forward and reverse shocks progress, the gas density behind the shocks increases rapidly, producing a thin shell of dense gas between the forward and reverse shocks \citep[][]{eladbrian,vlasov}. This shell is dense enough to be protected from radiation, and as the gas cools, molecules can begin to nucleate into dust particles. 
Because radio imaging shows that shocks are produced at the interface between the slow-moving toroidal outflow and the fast-moving spherical outflow, the dense shells (and dust formation) might be localized in the equatorial plane.

\begin{figure*}
\centering
\includegraphics[scale=1.0,angle=0]{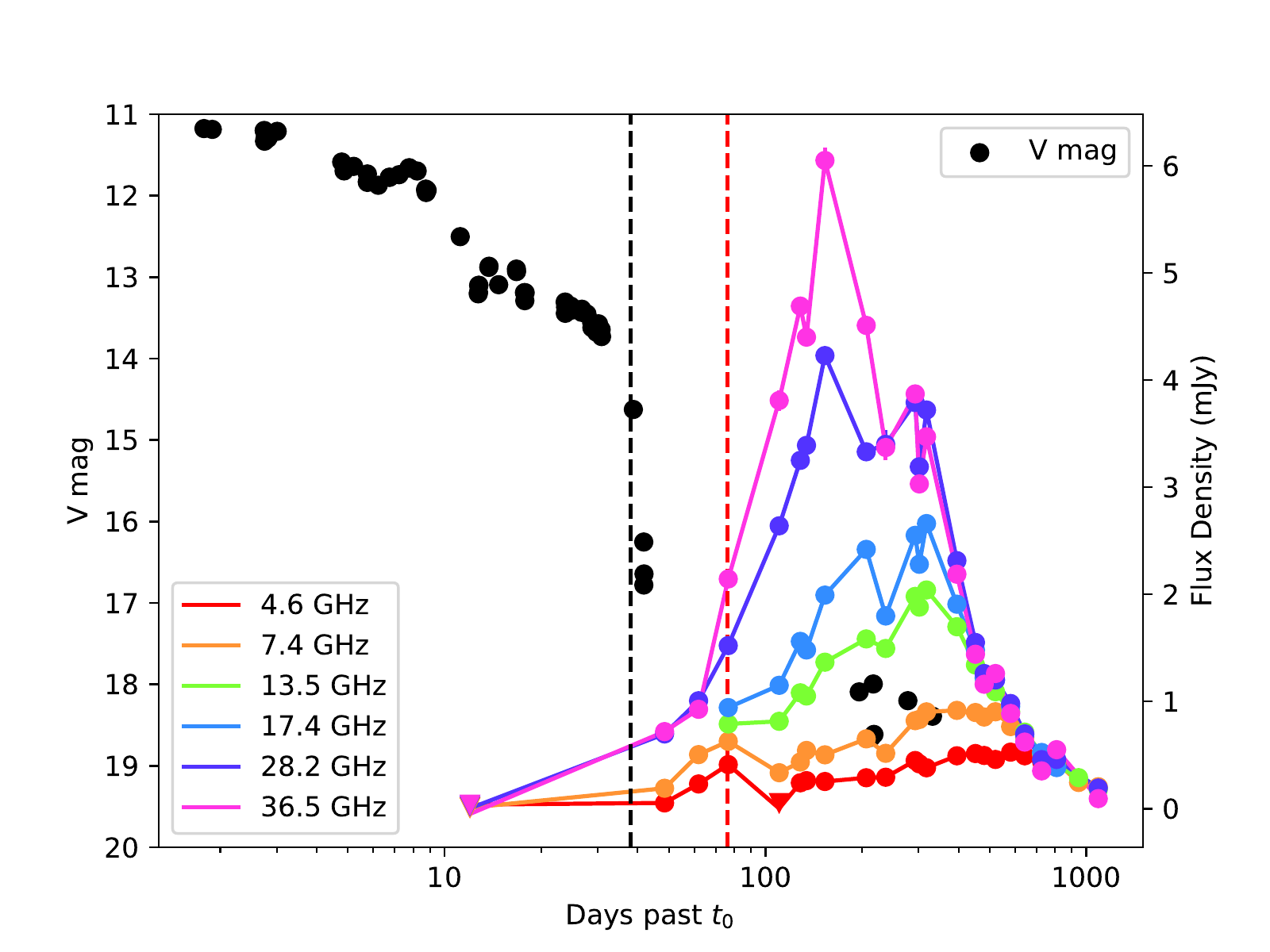}
\vspace{-5mm}
\caption{Optical and multi-frequency radio lightcurves of Nova V809 Cep as a function of days past 
discovery in the optical on 2013 February 2 ($t_0$). The triangles denote 
upper limits, and circles denote detections with significance of at least $3\sigma$. The high-frequency radio lightcurves peak between day 100 and day 300 (Epochs 5-12),
and there is a local maximum on day 76 (Epoch 4) at 
4.6 and 7.3 GHz (red vertical line).   Black points are V band magnitudes from \citep[][]{munari}. Error bars are smaller than the points.  The vertical black line denotes the onset of dust formation on day 37.}
\label{fig:fluxdata}
    \vspace{-5mm}
\end{figure*}

V809 Cep was first discovered in a nova eruption on 2013 February 2 by K.~Nishiyama and F.~Kabashima (CBET 3397), and we take this date to be the start of the eruption, $t_0$. V809 Cep reached peak optical brightness one day later \citep{munari}. Emission line profiles on day 220 and 352 for [OIII] 5007 $\AA$ and [NII] 6548, 6584 + H$\alpha$ had a double peak with a separation of 1160 km/s corresponding to a bulk velocity of 600 km/s \citep{munari}. Using the Echelle spectra and comparing with \citet[][]{brandblitz}, \citet[][]{munari} found the lower limit on the distance to be 6~kpc associated with the crossing of the Perseus or Outer arm. Additionally, using methods from \citet[][]{DD} and MMRD techniques, 
they refined their distance estimate to 6.5 kpc. We thus take the distance to be D=6.5 kpc.

In this paper, we use radio observations 
to show that nova V809 Cep produced non-thermal emission quasi-simultaneously with dust production. We postulate that the non-thermal emission occurred prior to the dust formation, but was initially absorbed by photo-ionized gas ahead of the shock.  In Section 2, we 
describe the technical details of the observations and data reduction. In Section 3, we present our results, and in Section 4 we discuss the implications of our findings in validating the shock-dust model. 

\section{Observations}

We monitored V809 Cep using the Karl G. Jansky Very Large Array Telescope (VLA) between 2013 February 14 and 2016 January 28 under programs 13A-455, 13B-057, and 15B-343. Between 2013 February and 2016 January, the VLA cycled through all configurations, and observed V809 Cep at C band (4 -- 8 GHz), Ku band (12 -- 18 GHz), and Ka band (26 -- 40 GHz). All the observations from 2013 February to 2015 April were performed in the 8-bit observing mode with 2-GHz bandwidth at all bands.  The observations in 2015 September and 2016 January were performed in the 3-bit mode with 4-GHz bandwidth at C band, 6-GHz bandwidth at Ku band, and 8-GHz bandwidth at Ka band. 
Table~\ref{tab:obslog} lists details for each observation. 

For observations carried out prior to 2013 May, we reduced the data using 
\texttt{AIPS} \citep{2003ASSL..285..109G}.  For observations carried out between 2013 May and 2016 May, we
reduced the data using standard routines 
in \texttt{CASA}  \citep{2007ASPC..376..127M}.To obtain radio flux densities, we fit
an elliptical Gaussian to the source image using \texttt{JMFIT} in \texttt{AIPS} and
\texttt{gaussfit} in \texttt{CASA}. 
We estimated uncertainties in the flux density by adding in quadrature the 
uncertainty from the fit with 
a systematic uncertainty from absolute flux calibration. At frequencies above 10 GHz, we took the absolute flux density uncertainty to be 
10$\%$, and at frequencies below 10 GHz,
we took the uncertainty introduced by the calibration to be
5$\%$. During each observation of V809 Cep, we also observed the source J2339+6010 and source 3C48. Source J2339+6010 was used for complex gain calibration and source 3C48 was used for absolute flux calibration. 

\begin{table*}
\centering
\begin{threeparttable}[t]
\caption{V809 Cep radio observations} 
\label{tab:obslog}
\begin{tabular}{p{0.12\linewidth}p{0.12\linewidth}p{0.12\linewidth}p{0.12\linewidth}p{0.12\linewidth}p{0.12\linewidth}p{0.12\linewidth}p{0.12\linewidth}}
\hline
Date & $t-t_0$\tnote{1} & MJD & Epoch & Configuration & Bands
Observed & Time (min)\tnote{2} \\
\hline
02/14/2013 & 12 & 56337.0 & 1 & D & C,Ka & 47.9\\
03/22/2013 & 48 & 56373.5 & 2 & D & C,Ka & 47.9 \\ 
04/04/2013 & 61 & 56386.9 & 3 & D & C,Ka & 47.9 \\ 
04/19/2013 & 76 & 56401.6 & 4 & D & C,Ka & 29.9 \\ 
05/23/2013 & 110 & 56435.4 & 5 & C & C,U,Ka & 29.9\\ 
06/10/2013 & 128 & 56453.5 & 6 & C & C,U,Ka & 29.9\\ 
06/16/2013 & 134 & 56459.3 & 7 & C & C,U,Ka & 29.9\\ 
07/05/2013 & 153 & 56478.3 & 8 & C & C,U,Ka & 29.9 \\ 
08/27/2013 & 206 & 56531.2 & 9 & C & C,U,Ka & 29.9\\ 
09/27/2013 & 237 & 56562.2 & 10 & B & C,U,Ka & 29.2 \\ 
11/22/2013 & 293 & 56618.0 & 11 & B & C,U,Ka & 29.9\\ 
11/30/2013 & 301 & 56626.8 & 12 & B & C,U,Ka & 29.9 \\ 
12/16/2013 & 317 & 56642.8 & 13 & B & C,U,Ka & 29.9\\ 
03/14/2014 & 405 & 56719.7 & 14 & A & C,U,Ka & 29.9 \\ 
04/29/2014 & 451 & 56776.5 & 15 & A & C,U,Ka & 30.2\\ 
05/28/2014 & 480 & 56805.6 & 16 & A & C,U,Ka & 30.2 \\ 
07/08/2014 & 521 & 56846.3 & 17 & D & C,U,Ka & 29.9 \\ 
09/06/2014 & 581 & 56906.2 & 18 & D & C,U,Ka & 29.9\\ 
11/07/2014 & 642 & 56968.0 & 19 & C & C,U,Ka & 29.9 \\ 
01/29/2015 & 726 & 57051.1 & 20 & B & C,U,Ka & 27.4\\ 
04/21/2015 & 808 & 57133.5 & 21 & B & C,U,Ka & 29.9\\ 
09/06/2015 & 946 & 57271.4 & 22 & A & Ka,U,S,C & 31.4 \\ 
01/28/2016 & 1090 & 57415.0 & 23 & C & Ka,U,S,C & 38.6\\
\hline
\end{tabular}
\begin{tablenotes}
    \item[1] We take the start time of the eruption, $t_0$, to be 2013 Feb 2.
    \item[2] Total time on source 
\end{tablenotes}
\end{threeparttable}
\end{table*}

\section{Results}
\begin{table*}
\begin{threeparttable}[t]
\caption{V809 Cep radio flux densities measured with the VLA
between 2013 Feb 14 to 2016 January 28, in mJy}
\label{tab:fluxdata}
\centering
\begin{tabular}{p{0.12\linewidth}p{0.12\linewidth}p{0.12\linewidth}p{0.12\linewidth}p{0.12\linewidth}p{0.12\linewidth}p{0.12\linewidth}}
\hline
Epoch & 4.6 GHz & 7.4 GHz & 13.5 GHz & 17.4 GHz & 28.2 GHz & 36.5 GHz\\
\hline
1 &  $<$0.051\tnote{1} & $<$0.018\tnote{1} & -- & --& $<$0.031\tnote{1}  & $<$ 0.039 \tnote{1} \\
2 & 0.052 $\pm$ 0.016 \tnote{2} & 0.189 $\pm$ 0.020 & -- & -- & 0.696 $\pm$ 0.040 & 0.718 $\pm$ 0.051\\
3 & 0.229 $\pm$ 0.027 & 0.503 $\pm$ 0.032 & -- & -- & 1.00 $\pm$ 0.028 & 0.926 $\pm$ 0.037\\
4 & 0.410 $\pm$ 0.061 & 0.627 $\pm$ 0.016 & 0.789 $\pm$ 0.015 & 0.942 $\pm$ 0.017 & 1.521 $\pm$ 0.031 & 2.143 $\pm$ 0.057\\
5 & $<$0.054 \tnote{1} & 0.334 $\pm$ 0.029 & 0.814 $\pm$ 0.025 & 1.150 $\pm$ 0.022 & 2.639 $\pm$ 0.065 & 3.809 $\pm$ 0.097\\
6 & 0.240 $\pm$ 0.027 & 0.434 $\pm$ 0.014 & 1.08 $\pm$ 0.015 & 1.56 $\pm$ 0.015 & 3.25 $\pm$ 0.039 & 4.69 $\pm$ 0.058\\
7 & 0.260 $\pm$ 0.033 & 0.542 $\pm$ 0.021 & 1.05 $\pm$ 0.017 & 1.48 $\pm$ 0.021 & 3.39 $\pm$ 0.0608 & 4.40 $\pm$ 0.082\\
8 & 0.252 $\pm$ 0.039 & 0.501 $\pm$ 0.015 & 1.366 $\pm$ 0.021 & 1.992 $\pm$ 0.024 & 4.228 $\pm$ 0.077 & 6.05 $\pm$ 0.12\\
9 & 0.287 $\pm$ 0.025 & 0.650 $\pm$ 0.024 & 1.583 $\pm$ 0.024 & 2.419 $\pm$ 0.035 & 3.33 $\pm$ 0.01 & 4.51 $\pm$ 0.03\\
10 & 0.292 $\pm$ 0.019 & 0.515 $\pm$ 0.030 & 1.494 $\pm$ 0.037 & 1.797 $\pm$ 0.089 & 3.40 $\pm$ 0.13 & 3.37 $\pm$ 0.12 \\ 
11 & 0.448 $\pm$ 0.018 & 0.820 $\pm$ 0.013 & 1.98 $\pm$ 0.02 & 2.55 $\pm $ 0.02 & 3.79 $\pm$ 0.05 & 3.87 $\pm$ 0.07 \\ 
12 & 0.417 $\pm$ 0.023 & 0.831 $\pm$ 0.015 & 1.88 $\pm$ 0.02 & 2.28 $\pm $ 0.02 & 3.19 $\pm$ 0.04 & 3.03 $\pm$ 0.05\\ 
13 & 0.380 $\pm$ 0.023 & 0.902 $\pm$ 0.016 & 2.04 $\pm$ 0.02 & 2.66 $\pm$ 0.02 & 3.72 $\pm$ 0.06 & 3.47 $\pm$ 0.08 \\ 
14 & 0.492 $\pm$ 0.017 & 0.916 $\pm$ 0.012 & 1.697 $\pm$ 0.013 & 1.907 $\pm$ 0.017 & 2.314 $\pm$ 0.040 & 2.187 $\pm$ 0.065 \\ 
15 & 0.513 $\pm$ 0.027 & 0.896 $\pm$ 0.014 & 1.34 $\pm$ 0.01 & 1.48 $\pm$ 0.02 & 1.55 $\pm$ 0.02 \tnote{5} & 1.44 $\pm$ 0.03 \\ 
16 & 0.495 $\pm$ 0.017 & 0.854 $\pm$ 0.011 & 1.22 $\pm$ 0.02 & 1.22 $\pm$ 0.02 & 1.26 $\pm$ 0.03 \tnote{5} & 1.16 $\pm$ 0.03 \\ 
17 & 0.458 $\pm$ 0.037 & 0.903 $\pm$ 0.029 & 1.09 $\pm$ 0.02 & 1.20 $\pm$ 0.03 & 1.20 $\pm$ 0.05 & 1.26 $\pm$ 0.08 \\ 
18 & 0.526 $\pm$ 0.033 & 0.765 $\pm$ 0.024 & 0.904 $\pm$ 0.020 & 0.952 $\pm$ 0.025 & 0.980 $\pm$ 0.062 & 0.887 $\pm$ 0.072 \\ 
19 & 0.492 $\pm$ 0.024 & 0.667 $\pm$ 0.014 & 0.713 $\pm$ 0.015 & 0.688 $\pm$ 0.017 & 0.697 $\pm$ 0.043 & 0.620 $\pm$ 0.051 \\ 
20 & 0.437 $\pm$ 0.034 & 0.503 $\pm$ 0.022 & 0.515 $\pm$ 0.022 & 0.523 $\pm$ 0.022 & 0.456 $\pm$ 0.044 & 0.351 $\pm$ 0.062\\
21 & 0.413 $\pm$ 0.020 & 0.432 $\pm$ 0.015 & 0.402 $\pm$ 0.015 & 0.382 $\pm$ 0.018 & 0.458 $\pm$ 0.042 & 0.549 $\pm$ 0.057 \\ 
22 & 0.265 $\pm$ 0.013 \tnote{3} & 0.243 $\pm$ 0.012 \tnote{4} & 0.288 $\pm$ 0.022 & 0.311 $\pm$ 0.032 \tnote{6} & 0.234 $\pm$ 0.079 \tnote{5} & 0.124 $\pm$ 0.071 \tnote{7} \\ 
23 & 0.305 $\pm$ 0.024 \tnote{3} & 0.203 $\pm$ 0.014 \tnote{4} & 0.176 $\pm$ 0.015 & 0.165 $\pm$ 0.011 \tnote{6} & 0.191 $\pm$ 0.024 \tnote{5} & 0.092 $\pm$ 0.025 \tnote{7} \\
\hline
\end{tabular}
\begin{tablenotes}
    \item[1] 
    We take the threshold for detection to be a flux density of at least three times the uncertainty ($3\sigma$). We list  non-detections as $2\sigma$. 
    upper limits. 
     \item[2] The flux density reported is measured at 4.7 GHz 
     \item[3] The flux density reported is measured at 5.0 GHz 
     \item[4] The flux density reported is measured at 7.0 GHz
     \item[5] The flux density reported is measured at 16.5 GHz
     \item[6] The flux density reported is measured at 29.5 GHz
     \item[7] The flux density reported is measured at 35.0 GHz
\end{tablenotes}
\end{threeparttable}
\end{table*}

At radio frequencies above 13 GHz, roughly speaking, the flux density rose over the course of five to ten months, and then faded, but at frequencies below 7.4 GHz, V809 Cep 
brightened and faded twice (Fig.~\ref{fig:fluxdata}). On 2013 March 22 (day 48-Epoch 2), five weeks after a non-detection on 2013 February 14 (day 12-Epoch 1), the VLA detected V809 Cep at 7.4, 28.2, and 36.5 GHz. 
The source continued to brighten and was detected at all observed frequencies (4.6, 7.4, 28.2, and 36.5 GHz) on 2013 April 19. On day 76 (Epoch 4), the brightness at 4.6 GHz and 7.4 GHz reached 
local maxima, 
at flux densities of 
$0.41 \pm 0.06$~mJy and 
$0.63 \pm 0.02$~mJy, respectively (see Figure~\ref{fig:fluxdata}). 
We refer to this low frequency brightening two to three months after the initial outburst as  the \textbf{early-time flare}. Even during the decay of the flare, the flux densities at 
high frequencies continued to rise and reached maxima between days 100 and 300 before beginning to decline 
(see 
Figure~\ref{fig:fluxdata}).
Table~\ref{tab:fluxdata} presents the flux densities and associated uncertainties. We 
take
a measured flux density of between three and five times the total uncertainty
to be a low-significance detection and 
a measured flux density of at least five times the total uncertainty to be a high-significance detection. Error bars correspond to 1$\sigma$ uncertainties. V809 Cep appeared as an unresolved point source at all epochs.

The radio spectrum evolved through four distinct stages: 1) a broken power law during the rise of the early-time flare; 2) a simple power law between epochs 5 and 
8 (Day 110 - 153), with a spectral index that 
remained stable at around $\alpha = 1.5$ to 1.6 for more than a month as the radio emission brightened again (taking $S_{\nu} \propto \nu ^{\alpha}$, where $S_\nu$ is the flux density at frequency $\nu$);
3) a broken power law again as the main decline started first at the highest and then lower frequencies; and 4) a simple, flat spectrum during the final decline.  Over the full three-year time scale of the eruption -- from 2013 March 22 (day 48) to 2016 January 28 (day 1090)-- the radio spectrum evolved from one 
in which flux density increased with frequency, as expected for optically thick emitting material, to one 
in which flux density decreased slightly with frequency.  We obtained the spectral index ($\alpha$) for a given date and frequency range by performing a linear regression using
\texttt{SciPy}'s least squares curve fitting method in log-log 
space, taking into account the uncertainties in the 
flux density measurements. We generally split the frequencies into low frequencies (4.6, 7.4 GHz), intermediate frequencies ( 13.5 GHz, 17.4 GHz), and high frequencies (28.2, 36.5 GHz). If the spectral indices at low, intermediate, and high frequencies were within 0.2 of each other, we fit them with 
a single powerlaw. If the spectral indices differed by more than 0.2, we fit the spectrum with a broken powerlaw. 

\begin{figure*}
\includegraphics[scale=0.7,angle=0]{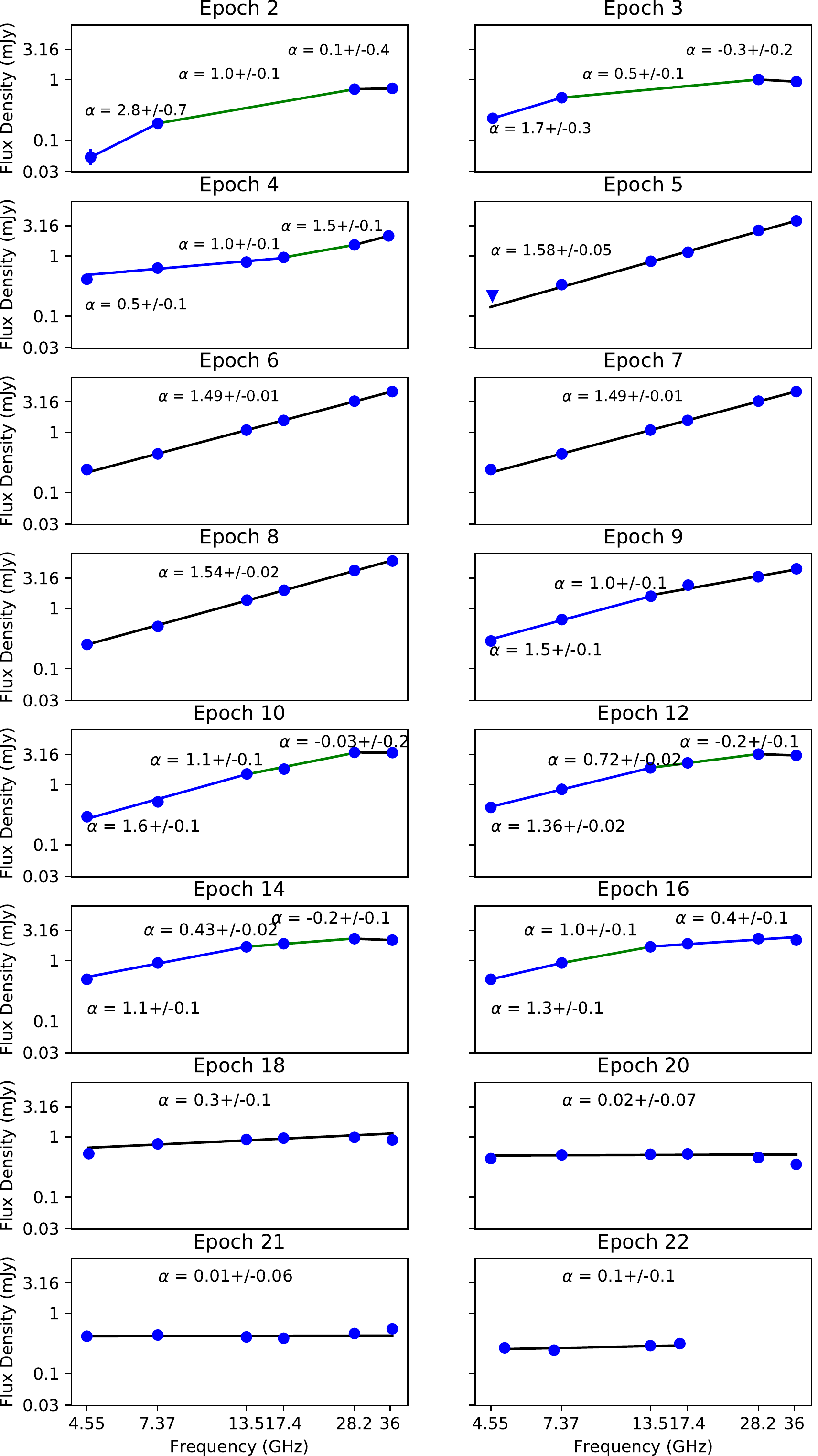}
\vspace{-1mm}
\caption{Evolution of the spectral energy distribution 
of V809 Cep for selected epochs between Epoch 1--23. Upper limits are plotted as triangles, and error-bars are 
typically smaller than the points. $\alpha$ represents the spectral index. The peak of the early-time flare occurred at Epoch 4.}
\label{fig:radiospectra}
    \vspace{-5mm}
\end{figure*}

Figure~\ref{fig:radiospectra} 
shows the radio spectral energy distribution 
for selected epochs between 2013 March 22 and 2016 January 28. During Epochs 2, 3, and 4 (stage 1), the spectral index at low frequencies (4.6 GHz and 7.4 GHz) 
differed from 
that at high frequencies. At low frequencies, the spectral index evolved from 
$\alpha = 2.8 \pm0.7$ in Epoch 2 to a much flatter $\alpha = 0.48 \pm 0.07$ in Epoch 4. In 
contrast, at high frequencies the spectral index evolved from being quite flat at Epoch 2 ($\alpha$ = 0.1 $\pm$ 0.4) 
to rising with $\alpha = 1.5 \pm 0.2$ at Epoch 4. 
During stage 3 (Epochs 10-16), the spectrum flattened first at the highest frequencies (Epoch 10) and subsequently at lower frequencies, leading to stage 4 (Epochs 18-22), 
during which time the spectrum was flat 
($\alpha$ = 0.1 $\pm$ 0.1 at Epoch 22) across the full range of frequencies. 

\section{Discussion}
\subsection{Early Time Radio Flare}
During the second and third month of the outburst, the nova produced a flare of low-frequency radio emission that brightened more quickly than expected for freely expanding thermal ejecta and reached a brightness that exceeded 
expectations for thermally emitting material
expanding at the speed of the principal optical absorbing system, suggesting that the flare was non-thermal in origin.
The first piece of evidence in support of non-thermal emission is the rapid brightening at low frequencies.  Between Epochs 2 and 3, the 4.6-GHz emission increased roughly as $t^6$, where $t$ here is the time since $t_0$.  Even if the material that generated the early-time flare was not ejected from the region of the white dwarf until four days after the start of the eruption \citep{munari}, the 4.6-GHz flux density increased faster than $t^4$.  During the same interval, the 7.4-GHz brightness increased as $t^4$.  Between Epochs 3 and 4, the 4.6-GHz emission continued to increase faster than $t^2$.  For a freely expanding blackbody with constant temperature, the brightness of thermal radio emission can only increase as fast as $t^2$, proportional to the increasing area of the source on the sky.  Although it would be possible for the size of a thermally emitting source to increase faster than $t^2$ if an ionization front was rapidly propagating through the ejecta, or if the electron temperature of the emitting material was increasing, the high radio brightness during Epochs 3 and 4 (see discussion below and Fig.~3) disfavors either of these possibilities.

%

The second piece of evidence in support of non-thermal emission is the strength of the low-frequency radio emission 
during Epochs 3 and 4. We parameterize the radio flux density in terms of a {\em brightness temperature, $T_b$} -- the minimum electron temperature required for the observed radio emission to arise from thermal Bremsstrahlung.
Brightness temperature is a measure of surface brightness.
If a uniform temperature, Bremsstrahlung-emitting medium is optically thick over its full area on the sky,
the brightness temperature 
corresponds to its physical temperature; however, if 
it is even partially optically thin, or if the radio emission emanates from a region that is smaller than the size of the putative spherical ejecta,
then the brightness temperature 
only provides a lower limit for the physical temperature. 
Assuming a spherically expanding shell, $T_b$ is given by: 
\begin{equation}
    T_{b}(\nu,t) = \frac{S_{\nu}(t)c^2D^2}{2 \pi k_{b} \nu^2 (v_{ej}t)^2},
\end{equation}
where c is the speed of light, $k_b$ is the Boltzmann constant, $t$ is the time since $t_0$, and  $v_{ej}$ is the maximum 
expansion velocity.  Using a distance of D = 6.5 kpc and a maximum ejection velocity of 1200 km/s \cite[corresponding roughly to the faster of the two principal absorption systems seen in optical spectroscopy;][]{munari},
the brightness temperature at 4.6 GHz reached $1.3 \pm 0.2 \times 10^5$ K (Epoch 4 (day 76); see Figure 3). 
For the early-time radio flare to have had a thermal origin would therefore require that on day 76 (Epoch 4), the ejecta included a shell of material with an electron temperature of at least $10^5$~K, that was at least as extended on the sky as the 1200~km/s flow, and that was dense enough to be optically thick at 4.6 GHz.  If the radio emission
emanated from knots, or a portion of the ejecta that subtended a smaller solid angle on the sky than the 
1200~km/s flow, the required electron temperature for the early-time flare to have been thermal is even higher. 
We argue below that such a large, hot, dense ejecta on day 76 (Epoch 4) seems unlikely.

Crucially, given that 
$T  \sim 10^4$~K is the approximate equilibrium temperature for a plasma cooled by forbidden emission lines \citep[][]{Westonaql}, 
photoionized ejecta from novae are typically expected to have electron temperatures of around this value.
In fact, \citet{2015ApJ...803...76C}
found that for a wide range of white dwarf masses, ejecta masses, and ejecta speeds, the temperature of the photoionized ejecta generally does not exceed a few times $10^4$~K at any point.  It is possible that the peak brightness of the early-time radio flare in V809~Cep could have been generated by the most spatially extended flow associated with the so-called diffuse enhanced optical spectroscopic system  \citep[which had speeds of up to 2000~km/s,][]{munari} if it had a temperature of $4 \times 10^4$~K and was completely optically thick. 
However, the low-frequency spectral index on day 76 (Epoch 4) was $\alpha = 0.5 \pm 0.1$, much lower than the $\alpha \approx 2$ expected for optically thick Bremsstrahlung, indicating that the radio-emitting ejecta were {\em not} optically thick at that time.
 
The expected temperature for photoionized ejecta and the
$\alpha = 0.5$ spectral index thus both argue against a thermal origin for the early-time flare.



\begin{figure*}
\centering
\includegraphics[scale=1.0,angle=0]{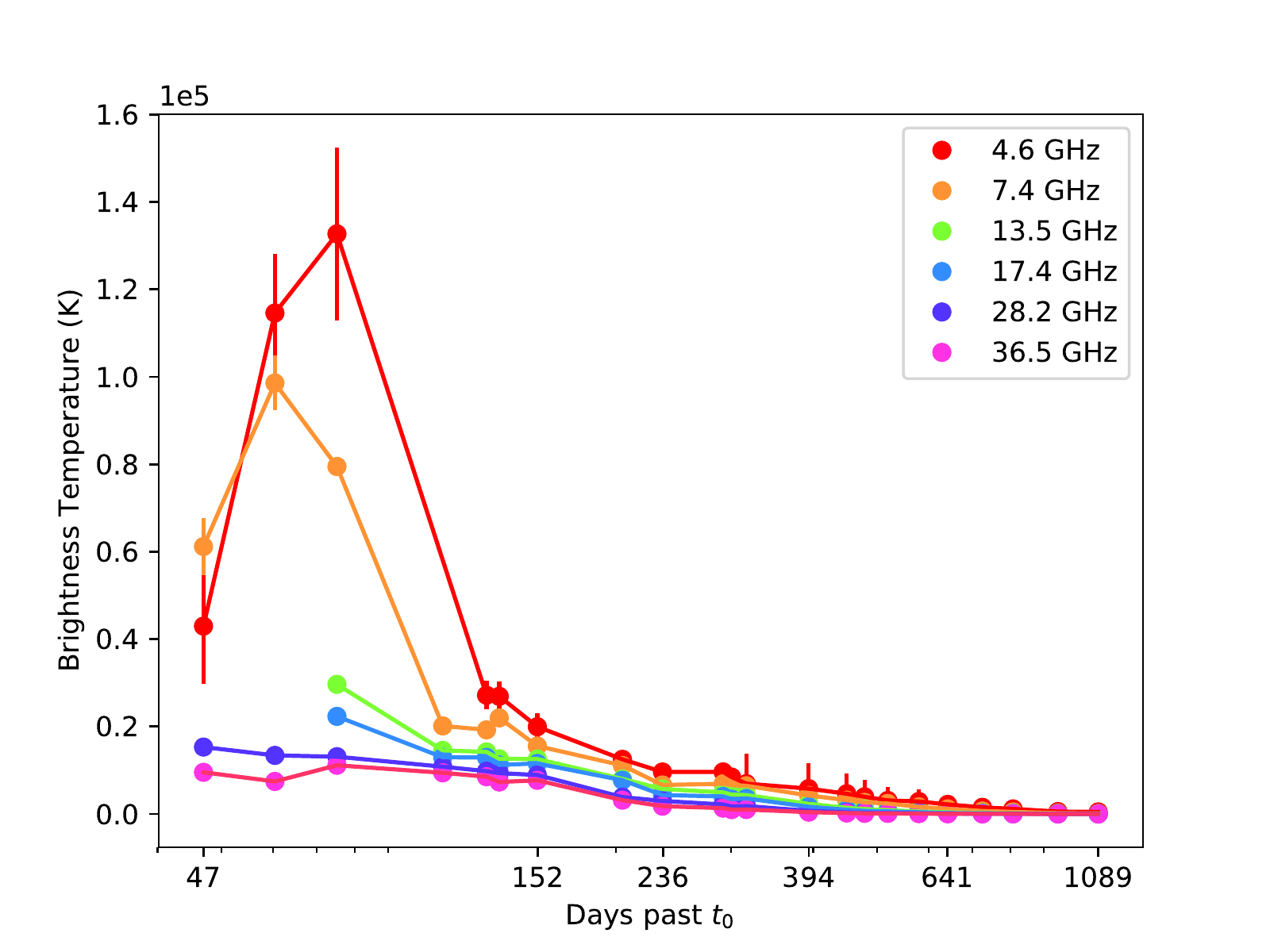}
\vspace{-2mm}
\caption{Brightness temperature as a function of time for each of the frequencies at which VLA observed V809 Cep using D=6.5kpc and source size described in text. The high brightness temperature in the first three months after eruption 
constitutes possible evidence for non-thermal emission and shocks}.
\label{fig:brightnesstemp}
    \vspace{-5mm}
\end{figure*}

The high $T_b$ is also unlikely to be due to a shell of shock-heated, $10^5$ K gas between the forward and reverse shocks.
\citet{eladbrian} explore the possibility of forward and reverse shocks in the ejecta producing a shell of cool ($\sim 10^4 K$), dense gas bounded by a shell of $T\sim 10^6$~K gas. 
We define $M_{hot,needed}$ as the mass of $10^5$~K gas that would be needed to make a shell of width $L_{cool}$ optically thick at radio frequencies and $M_{hot,expected}$ as the mass of $10^5$~K gas theoretically expected to be present. Below, we compare $M_{hot,expected}$ and $M_{hot,needed}$. 

\begin{figure}
\centering
\includegraphics[scale=0.6,angle=0]{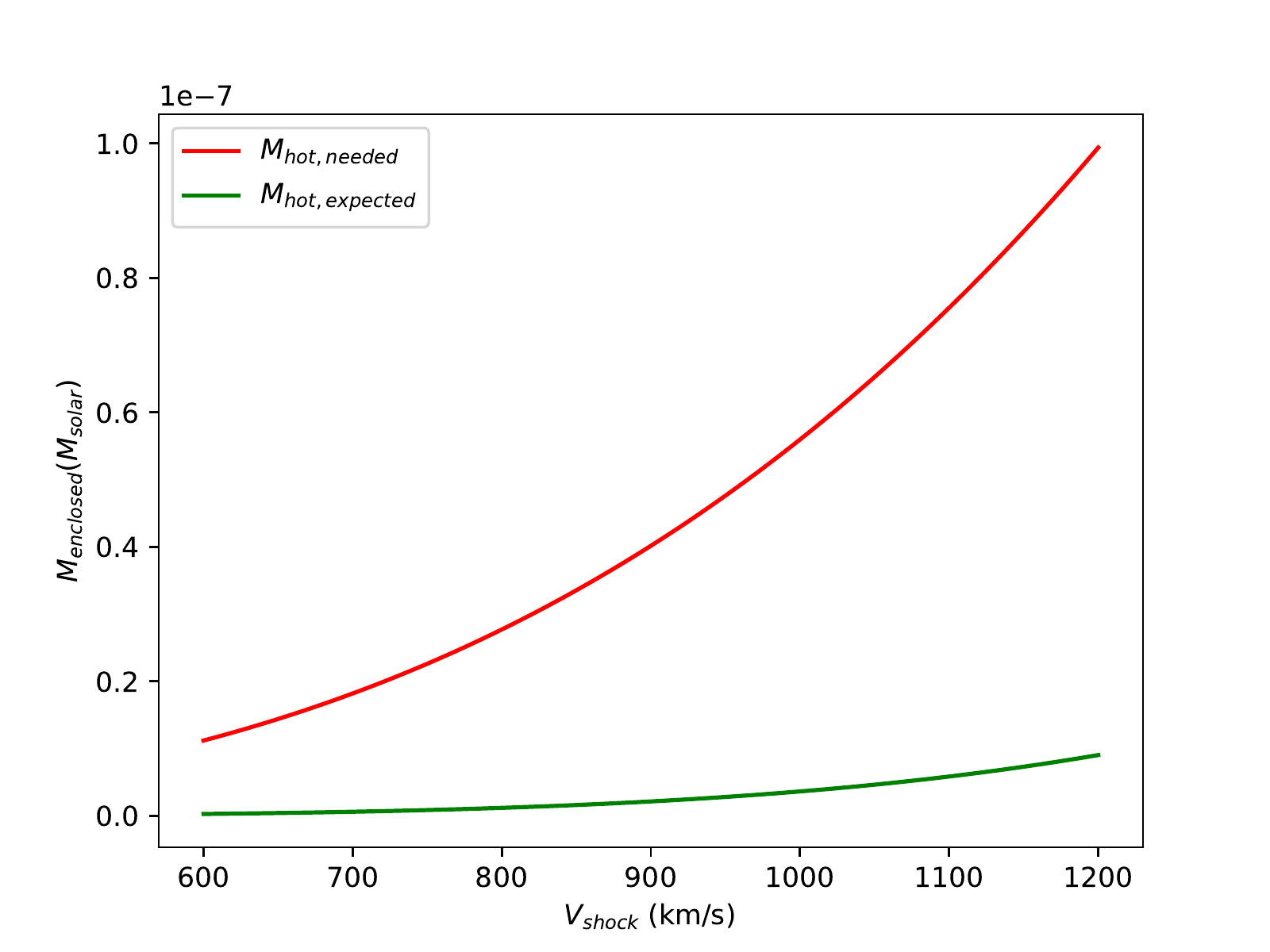}
\vspace{-2mm}
\caption{The mass expected in the $10^5$~K hot shell between the shocks ($M_{hot,expected}$) and the hot-shell mass 
needed to explain the early-time flare with just thermal emission ($M_{hot,needed}$) for a range of shock velocities from 600~km~s$^{-1}$ to 1200~km~s$^{-1}$.
}
\label{fig:mass_estimation}
    \vspace{-5mm}
\end{figure}

The expected geometrical thickness of the shell of hot gas can be approximated as the cooling length
\citep[$L_{cool}$;][]{eladbrian}:
\begin{equation}
    L_{cool} = v_{shock}t_{cool}/4.
\end{equation}
Because it is not 
clear whether the early-time radio flare
was the result of the forward or reverse shock, we take $v_{shock}$ to be the velocity of the forward shock. The reverse shock was directed into the fast moving, low density ejecta and 
thus had a higher velocity than the forward shock\, which was directed into higher density ejecta. This lower forward-shock velocity 
ends up being a more conservative estimate for $v_{shock}$ (see Fig.~4 for a comparison between $M_{hot,needed}$ and $M_{hot,expected}$ as a function of $v_{shock}$). The maximum 
velocity of the principle absorption system seen in optical spectroscopy was 1200 km/s, 
and the lowest expansion velocity observed was 600 km/s \citep[][]{munari}; therefore, we assume a shock velocity between 600 km/s and 1200 km/s.
As gas behind the shock cools radiatively,
$t_{cool}$ is given by \citep{eladbrian}: 
\begin{equation}
    t_{cool} = \frac{\bar{\mathbb{\mu}} (3/2) k_{b} T_{shock} }{\mu_{p} \mu_{e} n_{shock} \Lambda},
\end{equation}
where the mean molecular weight is $\bar{\mathbb{\mu}} = 0.74$ for a typical composition of a fully ionized classical nova
ejecta (\citealt{vlasov,2020MNRAS.497.2569S}, $\mu_{p} = 1.39$, and $\mu_{e} = 1.16$. From \citet{eladbrian}, we estimate the value of the cooling function ($\Lambda$) for $10^5$ K gas to be 
$10^{-20}$~cm$^3$/s. Assuming $\bar{\mathbb{\mu}} = 0.74$, the temperature of the shock, $T_{shock}$, is related to the shock velocity as \citep[][]{eladbrian}: 
\begin{equation}
    T_{shock} \approx 1.4 \times 10^7\;{\rm K} 
    \left(\frac{v_{shock}}{10^3\,{\rm km\; s}^{-1}}\right)^2.
\end{equation}
Multiplying the mean molecular weight by the expected shell thickness and number density at the location of the shell on day 76 -- for the appropriate range of shock velocities, a $1/r^2$ density profile (where $r$ is the distance from the white dwarf), a number density ($n_{shock}$) at $R_{shock} = v_{shock}t$ from the standard shock jump conditions \citep[][]{eladbrian} 
$n_{shock} = 4n$, and a total ejecta mass determined from the late-time radio evolution -- we find that $M_{hot,expected}$ likely fell between $10^{-10}$ and $10^{-8} M_{\odot}$
(see Fig.~\ref{fig:mass_estimation}).

%

For the early-time flare to have been due to thermal emission from
gas enclosed within $L_{cool}$, 
that gas must either have been optically thick with a temperature 
in excess of
$T_b = 1.3 \times 10^5$ K
or had an even higher temperature. In fact, the mass of hot gas that would have been needed to make the hot shell optically thick, $M_{hot,needed}$, is more than an order of magnitude greater than $M_{hot, expected}$. To estimate 
$M_{hot,needed}$, we assume the hot shell 
expanded with velocity $v_{shock}$. We derive the number density of electrons needed to make a $T_e \sim 10^5$~K shell optically thick at 4.55 GHz 
by setting $\tau_v = 1$ in the expression for the emission measure and taking a constant density profile within the hot shell.
The emission measure is $EM = \int_{r_o}^{r_i} n_{e}^2 dl$, where $n_{e}$ is the electron density. Following \citet{bodebook}, we calculate the emission measure needed to make the hot shell optically thick ($\tau_{\nu} = 1$) from: 
\begin{equation}
    \tau_{\nu} =1 = 8.235 \times 10^{-2}\left(\frac{T_e}{K}\right)^{-1.35}\left(\frac{\nu}{GHz}\right)^{-2.1}\frac{EM(\tau_{\nu})}{cm^{-6}pc} . 
\end{equation}
Figure.~\ref{fig:mass_estimation} 
shows both $M_{hot,expected}$ and $M_{hot,needed}$ enclosed in a shell of thickness $L_{cool}$ as a function of of shock velocity.

For $v_{shock}$ between 600 km/s to 1200km/s, $M_{hot,needed}$ exceeded $M_{hot,expected}$ by at least an order of magnitude. For $v_{shock}$ of 900km/s, $M_{hot,needed} = 1.5 x 10^{-8} M_{\odot}$ and $M_{hot,expected} = 8.5 x 10^{-10} M_{\odot}$. 
It is thus implausible for the high brightness temperature to have been the result of 
thermal emission from a shock-heated shell. This finding further supports our assertion that the high brightness temperature is most likely the result of non-thermal emission produced in a shock. 

Non-thermal emission from novae is widely accepted to be the result of internal shocks in the 
ejecta \citep[][]{laura_review,vlasov,v1324,Westonaql,fran,obrien}. The collision of the fast and slow flows results in shocks that are responsible for accelerating particles to relativistic speeds. 
Novae V1723 Aql \citep[][]{Westonaql}, V5589 Sgr \citep[][]{5589} and Nova 1324 Sco \citep[][]{v1324} had brightness temperatures of greater than $10^5$~K,
interpreted as evidence for radio synchrotron emission from shocks.  We 
similarly propose that the 
bright radio emission 
that the VLA detected on April 19 2013 (Epoch 4) 
was the result of shocks in ejecta from V809 Cep. V809 Cep thus joins six other classical novae 
with radio evidence for shocks: 
V906~Car \citep[][]{v906}, V1324~Sco \citep[][]{v1324},
V1723~Aql \citep[][]{Westonaql}, 5589~Sgr \citep[][]{5589}, V959~Mon
\citep[][]{chomui2014} and QU~Vul \citep[][]{QUvul}. 
Non-thermal radio emission 
has also been found in the eruptions of closely related objects
including the helium nova V445~Pup \citep{2021MNRAS.501.1394N} and 
the symbiotic system V407~Cyg \citep{2020A&A...638A.130G}. 

\subsection{Spectral Evolution} \label{sec:spectralevol}
The radio spectral evolution during the rise to maximum of the early-time flare in V809~Cep had some resemblance to theoretical predictions for synchrotron from internal shocks in novae by \citet[][]{vlasov}.
Those authors suggested that 
photo-ionized gas ahead of the shock initially absorbs radio synchrotron emission,
causing the spectral index ($\alpha$) to evolve from approximately 2.0, as 
expected for optically thick thermal emission, to 2.5 when the photo-ionized gas ahead of the shock becomes 
transparent and optically thick synchrotron emission begins to dominate the spectrum. \citet[][]{vlasov} predicted that the 
spectral index then gradually decreases toward the $\alpha = -0.7$ expected for
optically thin synchrotron
(the exact value depends on the energy
spectrum of the emitting particles, e.g. \citealt{1970ranp.book.....P}).  Later, the spectral index is expected to once again rise as the radio emission becomes dominated by optically thick thermal emission from the bulk of the expanding, photoionized ejecta. The peak of the early-time flare 
occurs as the synchrotron
transitions from optically thick 
to optically thin 
\citep[][]{vlasov}. 

At low frequencies (below 7.4 GHz), the radio emission during the 
early-time flare in V809~Cep was consistent with theoretical expectations -- if the VLA began detecting the flare when photoionized gas ahead of the shock had already become partially transparent. During the flare, the low-frequency
spectral index  transitioned from a high value suggestive of optically thick synchrotron emission (2.8 $\pm$ 0.7 in Epoch 2) to the more modest value of 0.5 $\pm$ 0.1 at the peak of the flare (during Epoch 4). Though the error bars on the low frequency spectral index at Epoch 2 are large, in combination with the high brightness temperature, the high spectral index implies optically thick synchroton emission. 

The 
flattening of the low-frequency spectral index at Epoch 4 suggests that 
the spectrum was indeed transitioning from one indicative of optically thick 
to optically thin synchrotron emission at that time, as predicted by \citet[][]{vlasov}. At Epoch 5, the low-frequency spectral index increased again, to 1.58 $\pm$ 0.05, as 
expected for optically thick thermal emission from the main, expanding ejecta beginning to dominate over the 
early-time synchrotron emission.

%
At high frequencies, the radio emission during the early-time flare can also be understood within the framework of \citet[][]{vlasov}.
Synchrotron emission from an expanding medium is expected to become optically thin, and the spectral index to decrease, at the highest frequencies first (29 GHz and above, in our case).  The high-frequency spectral indices at Epochs 2 and 3 of 0.1 $\pm$ 0.4 and -0.3 $\pm$ 0.2, respectively, thus follow the spectral evolution modeled in \citet[][]{vlasov} 
if the synchrotron emission had already become fairly 
optically thin at those frequencies by Epochs 2 and 3.
%
%
Moreover, because 
the high-frequency radio spectrum was already quite flat when the VLA first detected the source in Epoch 2, with the high-frequency flux densities well below an extrapolation of the low-frequency spectrum at that time,
synchrotron emission at frequencies higher than 7.4 GHz likely peaked before it peaked at low frequencies, and even prior to the first radio detection. Further supporting the possibility that any synchrotron peaked at high frequencies before low frequencies, at 7.4 GHz, the brightness temperature peaked during Epoch 3, an epoch before the peak at 4.6 GHz. 
Brightness temperature represents a lower limit on the physical temperature of an optically thin gas; therefore, we expect the brightness temperature to peak at high frequencies prior to peaking at low frequencies as the high frequencies become optically thin first. Figure~\ref{fig:brightnesstemp} displays a peak first a Epoch 3 (Day 61) at 7.4 GHz and later at Epoch 4 (Day 76) at 4.6 GHz. 
We thus posit that 
the early-time flare 
peaked at high frequencies prior to the first radio detection.   


Absorption by photo-ionized gas ahead of the shock  would initially obscure the non-thermal emission and produce a delay between the early-time radio maxima at progressively lower frequencies \citep[][]{vlasov}. As the photo-ionized gas ahead of the shock begins to diffuse and become optically thin at lower frequencies, the non-thermal emission from the shock becomes increasingly dominant first at higher frequencies and finally at lower frequencies.
So, hints that the synchrotron flare peaked at higher frequencies first (rather than at all frequencies simultaneously) suggest
that non-thermal emission from internal shocks in the V809 Cep ejecta was initially absorbed by photo-ionized gas ahead of the shock.
However, even if any synchrotron emission was initially somewhat absorbed by photoionized gas, the high spectral index of $2.8 \pm 0.7$ during Epoch 2 (Day 48) would indicate that we caught the flare when the putative synchrotron emission was already peeking through at low frequencies. 

The radio spectral evolution during Stage 2 (Epochs 5-9) further supports our assertion that the spectrum during stage 1 was attributable to 
an emission component that was physically distinct from the main photoionized ejecta.
As the early emission component thinned, the radio spectrum became increasingly dominated by optically thick thermal emission.
Indeed, during Epochs 4 (Day 76), the spectral index at high frequencies rose to 1.5 $\pm$ 0.1, indicative of optically thick thermal emission. During Epoch 5 through 9 (Day 110 - 206) , the 
spectral index for all frequencies increased
to 1.58 $\pm$ 0.05 and remained 
around 1.5 as the ejecta expanded and the radio emission brightened for the remainder of stage 2. Although the 
stage 2 spectral indices were lower than the expected 2.0 for optically thick thermal emission from a medium with an infinitely sharp outer edge, values 
of 1.5 are typical for actual nova remnants. Therefore, the radio spectral evolution was consistent with emission from an increasingly optically thin
synchrotron source and a gradually brightening optically thick thermal component. 
During stage 3 and 4, the radio spectrum was typical of a thermal source 
becoming increasingly optically thin, with the spectrum flattening at progressively lower frequencies as gas began to become optically thin first at high frequencies ($>$ 29 GHz) and by Epoch 20 (Day 726) at all frequencies. 


\begin{center}
\captionof{table}{V809 Cep timeline} 
\label{tab:timeline}
\begin{tabular}{ |l|l|c| } 
 \hline
 Event & Date & $t-t_0$ \\ 
 \hline
 Optical Detection & 2013 February 2 & 0 \\
 Optical Peak & 2013 February 3 & 1 \\ 
 Onset of dust production & 2013 March 11 & 37 \\ 
 Synchrotron detected & 2013 March 22 & 48 \\ 
 Peak of Low-Frequency synchrotron & 2013 April 19 & 76 \\
 \hline
\end{tabular}
\end{center}

\subsection{Quasi-simultaneity of synchrotron flare and dust formation}
Around the 
start of the early-time flare,
dust production also commenced, providing evidence for a connection between shocks and dust formation. 
In the 
shock-dust model \citep[][]{andrea}, internal shocks in the 
ejecta 
lead to dust condensation. We would naively expect, therefore, to see evidence for the shocks {\em prior} to the onset of dust formation, which began around 2013 March 11 (day 37; \citealt{munari}).
The VLA did not observe V809 Cep between day 12 (Epoch 1) and day 48 (Epoch 2) of the eruption,
so we cannot pinpoint the exact time at which synchrotron emission appeared (or peaked at high frequencies). 
But on day 48 (Epoch 2), 11 days past the onset of dust formation, we detected radio emission almost certainly associated with shocks within the ejecta.
%
%
Therefore, both 
signatures of shocks and dust appeared during the one month gap between Epoch 1 (Day 12) and Epoch 2 (Day 48). 
Given the evidence that the early-time flare did not peak at the same time at all frequencies and therefore that any synchrotron was initially obscured, the radio observations are consistent with the shocks arising either slightly before or around the same time as the onset of dust production.


%


\subsection{Properties of the Thermal Ejecta}

Starting about nine months into the eruption, the radio emission was reasonably consistent with Bremsstrahlung (free-free emission) from freely expanding, photoionized ejecta.  We modeled the main ejecta as a Hubble-flow  
%
\citep[][]{gehrz_2008,v1324,Westonaql,hubbleflow1,hubbleflow2}, 
in which  the ejecta 
consists of a homologously expanding isothermal spherical shell of gas
bound between inner and outer radii $r_{inner}$ and $r_{outer}$, with a $r^{-2}$ density profile (where $r$ is the distance from the white dwarf), and a volume filling factor, $f$. 
The filling factor is a measure of inhomogeneities in the 
ejecta, 
such as clumps \citep[e.g.,][]{Westonaql,1996ASPC...93..174H}.  
We assume no inhomogeneities ($f = 1$).
If the ejecta were actually clumpy, that
would reduce the ejecta mass needed to generate the observed radio emission \citep[][]{Westonaql,1996ASPC...93..174H,clump1,clump2,clump3}. 

At low frequencies ($<$ 7.4 GHz) during the first six or so months of the outburst, the model fit the observed flux densities poorly, and as expected, was unable to account for the early-time flare.  
At 13.5 and 17.0 GHz, the discrepancy between the model and the observed flux densities 
during the first few months of the eruption supports our contention (in Sec.~\ref{sec:spectralevol}) that the flare at these frequencies could have peaked 
before Epoch 4 (Day 76). At these frequencies, the Hubble-Flow model matched the observed flux densities well 
after Epoch 4 (Day 76). At Epoch 4 (Day 76), however, the observed flux density at 17.4 GHz was 
$0.94 \pm 0.02$~mJy, 0.61~mJy in excess of the
the model flux density 
of 0.33~mJy. At 13.5 GHz, the observed flux density was 
$0.789 \pm 0.002$, 
0.58 mJy in excess of the model flux density. 
That the observed 13.5- and 17.4-GHz flux densities were far in excess of the model during Epoch 4 (Day 76)
indicates that any synchrotron emission was still dominant at 
those frequencies at that time and could therefore have peaked 
even earlier (the VLA did not collect data at these frequencies during Epochs 2 and 3 (Day 48 and Day 61)).

\begin{figure*}
\centering
\includegraphics[scale=1.0,angle=0]{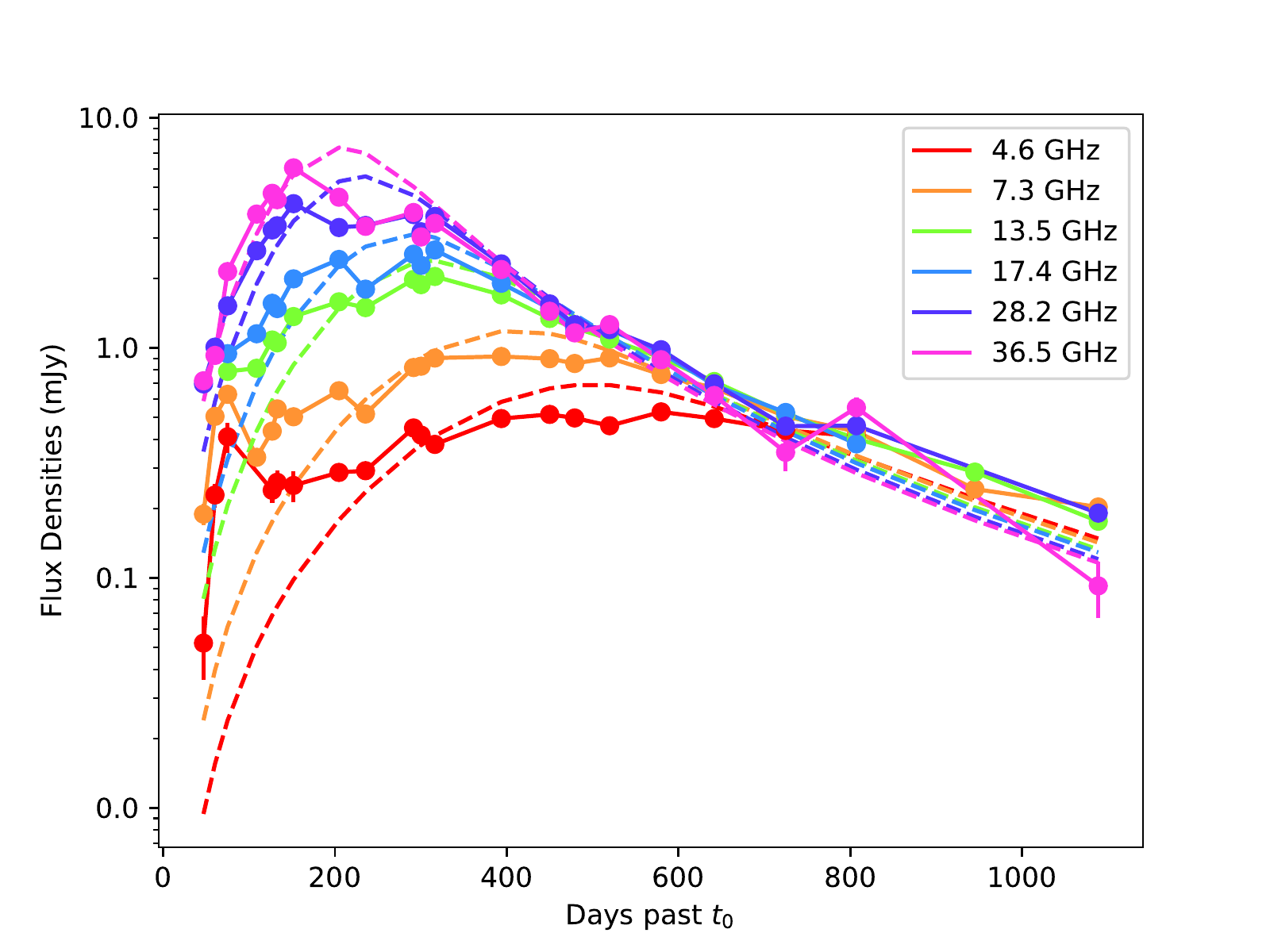}
\vspace{-5mm}
\caption{Hubble Flow model 
compared with observed flux densities. The dashed lines show the model flux densities, and the solid points denote the observed flux densities. The model fluxes are calculated using $M_{ejected} = 1.1 \times 10^{-4} M_{\odot}$, $T = 0.8\times 10^4 K$, $v_{ej}$ = 1200 km/s,$v_{min}$=600km/s, and D = 6.5 kpc}
\label{fig:fluxdataandmodel}
    \vspace{-5mm}
\end{figure*}

At 28.8 and 36.5 GHz, and after approximately days 200-300 (>Epoch 9) at all frequencies, however, the model 
reproduces the observed flux densities moderately well. 
Using $D = 6.5$~kpc, $v_{ej}=1200$~km~s$^{-1}$, and $v_{min}=600$~km~s$^{-1}$, we determined the best-fit ejecta mass and temperature by employing the Markov Chain Monte Carlo program pymc.
We found $M_{ejected} = (1.1 \pm 0.1) \times 10^{-4} M_{\odot}$ and $T = (0.8\pm0.2)\times 10^4 K$ when the hubble flow model was fit to the reduced flux densities starting from Epoch 2 (Day 48) (first detection).  Figure~\ref{fig:fluxdataandmodel} displays the observed flux densities as well as the modeled free-free emission for $D=6.5$~kpc, $M_{ejected} = (1.1 \pm 0.1) \times 10^{-4} M_{\odot}$ and $T = (0.8\pm0.2)\times 10^4$~K.

\section{Conclusions}
\begin{itemize}
  \item The radio light-curves at 4.6 GHz and 7.4 GHz display an early-time flare 
  that peaks at Epoch 4 (day 76). At 4.6 GHz, the brightness temperature 
  at that time was at least 1.5 $\times 10^5 K$. 
  \item At low frequencies ($<$ 7.4 GHz) the rise of the early-time flare began at Epoch 2 (Day 48, 2013 March 22). The spectral index at low frequencies at Epoch 2 (Day 48) was 2.82 $\pm$ 0.71.
  \item The high brightness temperature at the peak of the early-time flare at Epoch 4 (Day 76) combined with the high spectral index at Epoch 2 (Day 48) suggests the presence of non-thermal emission due to internal shocks in the nova ejecta. 
  This is the first report showing evidence for shocks in V809 Cep. 
  \item The discovery of non-thermal radio emission early in the evolution of the eruption puts V809 Cep in the same category as six other novae known to show radio evidence for shocks: V906 Car \citep[][]{v906}, V1324 Sco  \citep[][]{v1324}, V1723 Aql \citep[][]{Westonaql}, V5589 Sgr \citep[][]{5589}, V959 Mon \citep[][]{chomui2014},and QU Vul \citep[][]{QUvul}. 
  \item At late times (after Epoch 4 (Day 76)), the radio 
  emission was consistent with thermal emission from freely expanding ejecta with a maximum ejecta speed of 1200km/s, $v_{min}$ = 600km/s, D =6.5 kpc, $T = 0.8\times 10^4 \times 10^4 K$, and a $M_{ejected} = 1.1 \times 10^{-4} M_{\odot}$. 
  \item Our finding that a radio synchrotron flare associated with internal shocks in the ejecta began during a roughly month-long period of time during which dust formation also commenced supports the idea that shocks could form ideal environments for dust formation around novae. 
  
  \item If shocks in the nova ejecta lead to dust formation, the 11 day delay between the onset of dust and the rise of the early-time flare suggests that the radio emission was absorbed by photo-ionized gas ahead of the shock. 
\end{itemize}

\section*{Data availability}
The data underlying this article will be shared on reasonable request to the corresponding author. The uncalibrated radio observations are publicly available from the NRAO Data Archive (archive.nrao.edu or data.nrao.edu)

\section*{Acknowledgements}
The National Radio Astronomy Observatory is a facility of the National Science Foundation operated under cooperative agreement by Associated Universities, Inc.  A.~B. and J.~L.~S. acknowledge support from NSF grant AST-1816100 and  Heising-Simons Foundation grant 2017-246, as well as NRAO 362918. Laura Chomiuk, Adam Kawash, Elias Aydi, and Kirill V. Sokolovsky were supported by NSF grant AST-1751874 and a Cottrell fellowship of the Research Corporation.

\bibliographystyle{mnras}
\bibliography{main}

\end{document}